\documentclass[letterpaper]{nearbeam}
\usepackage[dvips]{graphics}
\input epsf
\newcommand{\eg}{{\em e.g.\ }}
\newcommand{\ie}{{\em i.e.\ }}

\begin{document}
\title 
{\rightline{\rm\normalsize IIT-HEP-98/1}
\bf BTeV/C0\thanks{Invited talk presented at the {\sl International Symposium 
on Near Beam Physics}, Fermilab, Sept.\ 22--24, 1997.}}
\author{Daniel M. Kaplan\thanks{Email: kaplan@fnal.gov}
     \\ {\sl Illinois Institute of Technology, Chicago, IL 60616, USA} 
     \\
 for the BTeV Collaboration         
}

\maketitle

\begin{abstract}
The physics goals and techniques of the proposed BTeV experiment at the C0 
Tevatron interaction area are summarized, with emphasis on aspects of the 
experiment that depend on near-beam issues. BTeV aims to carry out a 
comprehensive study of rare processes (especially {\em CP} violation) in charm 
and beauty decay starting in collider Run II. Vertex detectors will be deployed 
within a few mm of the beam. Early running may employ a wire target in the beam 
halo.
\end{abstract}

\section{Introduction}

The BTeV  collaboration is proposing to carry out a dedicated heavy-quark 
collider experiment in the C0 interaction region at the Tevatron. The main 
goals of BTeV  are to search for {\em CP} violation, mixing, and rare 
flavor-changing neutral-current (FCNC) decays of $b$- and $c$-quark hadrons at 
unprecedented levels of sensitivity. Each year of BTeV collider operation is 
expected to produce ${\cal O}(10^{11}$) $b$ hadrons and ${\cal O}(10^{12}$) $c$ 
hadrons, to be compared with ${\cal O}(10^{7}$) of each available at the 
$e^+e^-$ ``$B$ Factories" and ${\cal O}(10^{9}$) $b$ events per year at the 
HERA-$B$ fixed-target experiment. The BTeV spectrometer is being designed 
to make optimal use of the produced samples, avoiding many of the compromises 
necessary in general-purpose detectors.

The rationale for sensitive $b$-quark studies has been discussed 
extensively~\cite{BCP}-\cite{CP-review}. In a nutshell, the goal is to test
thoroughly the
Kobayashi-Maskawa (KM)~\cite{KM} mechanism -- the Standard-Model explanation 
for {\em CP} violation -- in a regime in which large effects are expected, as 
opposed to the ${\cal O}(10^{-3}$) effects observed in the $K^0$ 
sector~\cite{K0CP,CP-review}.  
The KM model, while compatible with all known experimental evidence, is not 
unique, and it is appropriate to regard the origin of {\em CP} violation as a 
key unsolved 
problem of contemporary science. The baryon asymmetry of the 
universe leads us to think~\cite{cosmic-baryon} that {\em CP} violation beyond 
that
predicted in the KM model should exist~\cite{Quinn}. 
The over-arching question in particle physics today is, what ``new physics" 
underlies the Standard Model?\footnote{The Standard Model, while consistent 
with all established experimental results, has more than twenty free parameters 
(the lepton masses, quark masses and mixing parameters, coupling constants, 
Weinberg angle, 
Higgs mass, etc.)\ and thus is generally considered to be 
only an approximation. New physical effect(s) yet to be discovered
are presumed to determine the values of these parameters.}
It is possible that $K^0$
{\em CP} violation arises in part or even entirely from
physics outside the Standard Model, in which case
it is the only 
new-physics signature that has already been observed.

Many experiments now seek to address this topic. The $B$-Factory and HERA-$B$ 
groups are vying to be the first to observe {\em CP} violation in $B$ decay, 
and the 
CDF and D0 groups are not far behind. However,
it is likely that these efforts, while adequate to observe effects, will not 
suffice for the thorough investigation that the importance of the topic demands.

High-sensitivity charm studies are complementary to beauty studies. In the
Standard Model, {\em CP} violation, mixing, and FCNC decays, all relatively
large in beauty, are drastically suppressed in charm~\cite{Burdman}. Any
contribution from new physics will thus stand out dramatically. 
For example, new physics might be Higgs-like and couple to quark
mass~\cite{Bigi87}, or might couple
more strongly to ``up-type"\footnote{{\em i.e.}, the $u$, $c$, and $t$ quarks}
than ``down-type"\footnote{{\em i.e.}, the $d$, $s$, and $b$ quarks}
quarks~\cite{up-down}. 
In such scenarios, charm has the biggest new-physics signal-to-background 
ratio of any quark.
On the experimental side one has (compared to beauty)
large production cross sections, large branching ratios to final states of
interest, and straightforward tagging via the $D^{*+}\to D^0\pi^+$ decay chain.
The experimental approach taken by BTeV, featuring a primary trigger based on
the presence of secondary vertices, naturally provides high charm and beauty
sensitivity simultaneously. We can thus carry out a ``two-pronged assault" on
the Standard Model.

\section{Standard-Model {\em CP} Violation}

\subsection{The CKM Quark-Mixing Matrix}

The KM mechanism for {\em CP} violation invokes a non-zero phase in the
Cabibbo-Kobayashi-Maskawa (CKM) quark mixing matrix~\cite{Cabibbo,KM}, $$  V =
\left( \begin{array}{ccc}              V_{ud} & V_{us} & V_{ub} \\             
V_{cd} & V_{cs} & V_{cb} \\              V_{td} & V_{ts} & V_{tb} \end{array}
\right)\,. $$ The matrix $V$ parametrizes the coupling of the $W$ bosons to the
quarks
in a way that allows the generations to mix. For example, instead of coupling 
the $u$ quark only to the $d$, $W^+$ emission couples the $u$ to the linear 
combination $$V_{ud}|d\rangle+V_{us}|s\rangle+V_{ub}|b\rangle\,,$$ with similar
expressions for the couplings to the 
$c$ and $t$ quarks. This generation mixing provides an explanation for the 
observed non-stability of the $s$ and $b$  quarks. 

As is well known, for two generations of quarks, the quark mixing matrix is 
real and has one free parameter, the Cabibbo angle~\cite{Cabibbo}. Being 
unitary, for three quark generations the matrix depends on only four 
independent parameters, including one non-trivial phase~\cite{KM}. Certain 
decays can occur via more than one Feynman diagram in such a way that the 
interference term between the diagrams contains this phase. When the decay 
width for such a reaction is compared to that for the {\em CP}-conjugate 
reaction, 
the dependence on the CKM phase (whose sign changes under {\em CP}) can result 
in a 
{\em CP} asymmetry, \eg
$$
A\equiv \frac{\Gamma(B\to f)-\Gamma(\overline{B}\to \overline{f})}
{\Gamma(B\to f)+\Gamma(\overline{B}\to \overline{f})}\ne 0\,,
$$
which will depend on the decay time if the interference involves 
$B\overline{B}$ mixing.

\begin{figure}
\centerline{\epsfxsize=3.25 in \epsffile{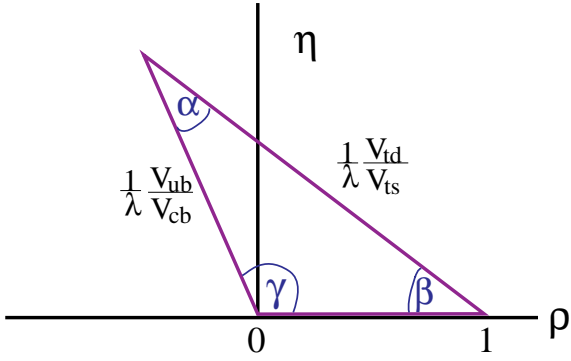}}
\caption{The ``unitarity triangle" for couplings of the $b$ 
quark, expressed in terms of the $\lambda$, $\rho$, and $\eta$ variables of the
Wolfenstein parametrization~\cite{Wolfenstein} of the CKM
matrix.\label{fig:triangle}}
\end{figure}

The unitarity of the CKM matrix further implies that the product of any two of 
its
rows or columns be zero. One such relationship is
$$
V_{ud}V_{ub}^*+V_{cd}V_{cb}^*+V_{td}V_{tb}^*=0\,.
$$
This relationship constrains mixing rates and {\em CP} asymmetries in various 
decays 
of beauty hadrons. Since it states that three complex numbers sum to zero, it 
can be visualized as defining a triangle in the complex plane 
(Fig.~\ref{fig:triangle}). Because (unlike the case in the $K^0$ and charm 
sectors) the sides of this triangle are all roughly similar in 
length~\cite{CP-update}
(Fig.~\ref{fig:triangle2}), the 
angles are expected to be large. Since the angles determine the {\em CP} 
asymmetries, these should be uniquely large in beauty decays.

\subsection{Studying the Unitarity Triangle}

\begin{figure}
\centerline{\epsfxsize=3.25 in \epsffile{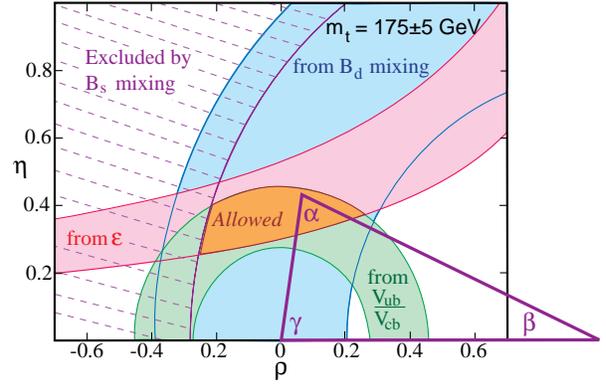}}
\caption{Current knowledge of the CKM triangle, based on experimental 
constraints on the lengths of its sides from $B$ decays, and on the position of 
its apex from the $\varepsilon$ parameter of $K^0$ {\em CP}
violation, with estimated $1\sigma$ error bands.
\label{fig:triangle2}}
\end{figure}

The task of verifying the KM model then reduces to measuring 
enough of the mixing and asymmetry parameters to prove that the triangle is 
indeed closed, \ie that its angles and the lengths of its sides are 
consistent. In addition it must satisfy constraints from {\em CP} violation in 
the 
$K^0$ sector (Fig.~\ref{fig:triangle2}). Ideally one would make enough 
different measurements to verify 
that {\em all} decays constrained by the unitarity triangle satisfy the 
constraints. This task is made difficult by the small 
branching ratios for 
interesting $B$-hadron final states (\eg $1.7\times10^{-5}$ for $B_d\to J/\psi 
K_S\to\mu^+\mu^-\pi^+\pi^-$), thus a large $b\bar b$ production cross section 
is required. Since $\sigma_{b\bar b}\sim 100\,\mu$b at $\sqrt{s}=2$\,TeV,
the Tevatron collider is a natural venue for such studies.

The angle $\beta$ can be determined from the {\em CP} asymmetry in $B_d\to 
J/\psi K_S$ with essentially no theoretical uncertainty. Since this mode also
has a clean
experimental signature in the $J/\psi\to\,$dileptons decay and (compared to
other modes with large expected {\em CP} asymmetries) a relatively large 
compound
branching ratio, it is sometimes called the ``golden" mode for $B$ {\em CP}
violation. Its {\em CP} asymmetry is expected~\cite{CP-update} to be
$\sim$0.5 in the Standard Model
and is likely to be measured by $\approx$2002 in the next round of
experiments. 

The other two angles of the unitarity triangle are considerably harder to 
determine. It is often stated that $\alpha$ is measured in
$B_d\to\pi^+\pi^-$. The measurement suffers from significant drawbacks.
First, the branching ratio is small ($<1.5\times10^{-5}$ at 90\%
C.L.~\cite{CLEO-hh})
and has yet to be definitively
established. Second, the larger branching ratio observed for
$B_d\to K^+\pi^-$~\cite{CLEO-hh} 
imposes stringent experimental requirements on
hadron identification and mass resolution to allow adequate suppression of
$K\pi$ background, and also
implies a significant contribution to
$B\!R(B_d\to\pi^+\pi^-)$ from penguin diagrams, whose {\em CP} asymmetry is 
difficult to relate to CKM angles. 
Nevertheless, the measurement of the {\em CP} asymmetry in
$B_d\to\pi^+\pi^-$ will be an important step forward and will furnish a
significant constraint on models of {\em CP} violation.

Various methods of determining $\gamma$ have been discussed and have various
advantages and
drawbacks. A promising method appears to be comparison of branching
ratios for $B^+\to {}^{^(}\overline{D^0}{}^{^)} K^+$ and $B^-\to
{}^{^(}\overline{D^0}{}^{^)} K^-$~\cite{Dunietz}.
Both of these can occur via two processes that interfere, namely
$B^+\to \overline{D^0} K^+,\,\overline{D^0}\to K^+\pi^-$ and $B^+\to
D^0 K^+,\,D^0\to K^+\pi^-$ (and charge-conjugates). 
Since the first proceeds via
$b\to u$ conversion while the second includes a doubly Cabibbo-suppressed $D^0$
decay, both are highly suppressed processes, leading to the favorable situation
where the interference between them can have a relatively large effect (of 
order unity) on
branching ratios. On the other hand, the branching ratios for these modes are 
expected to be ${\cal O}(10^{-6})$. Another method is via the
mixing-induced {\em CP} asymmetry in $B_s (\overline{B_s})\to D_s^\mp K^\pm$;
this measurement will require excellent decay-time resolution given the rapid
expected $B_s\overline{B_s}$-mixing oscillations.

We see that a complete test of the KM model will require very large $B$ 
samples. Only hadroproduction can supply such 
large numbers of events. Furthermore, since several of the decay modes of 
primary interest are to all-hadronic final states, a significant physics 
penalty is paid if the typical $B$ trigger, requiring high-$p_t$ leptons from 
semileptonic or $B\to J/\psi$ decays, is employed. We are thus led to the BTeV 
strategy: a first-level trigger based on decay-vertex reconstruction.

BTeV's sensitivity has been estimated~\cite{eoi} as $\pm$0.04 in $\sin{2\beta}$
and
(ignoring penguin contributions) $\pm$0.1 in $\sin{2\alpha}$. These
are for one year of running at the nominal luminosity of
$5\times10^{31}$\,cm$^{-2}$s$^{-1}$. We are investigating our sensitivity to 
$\gamma$ and also the possibility of running at higher luminosity.

\section{Non-Standard-Model {\em CP} Violation}

A variety of extensions to the Standard Model (SM) have been considered in
which
{\em CP}-violating phases can arise elsewhere than in the CKM matrix. Possible 
non-Standard sources for {\em CP} violation include additional Higgs doublets, 
non-minimal supersymmetry, massive $W$'s with right-handed couplings 
(``left-right-symmetric" models), leptoquarks, a fourth generation, 
etc.~\cite{CP-review,CP-survey}. Such mechanisms could be responsible for all
or part of $K^0$ {\em CP} violation.

These models have various attractive features. For example, an enlarged Higgs 
sector is a relatively natural and straightforward extension of the SM, 
especially since we know of no reason (other than Occam's Razor!)\ why, 
assuming Nature opted to implement the Higgs mechanism, she should have 
stopped after only one physical Higgs boson. Left-right-symmetric models are 
appealing in that they provide a unified explanation for both parity and {\em 
CP} violation. And in such extensions of the SM, the CKM phase could be 
exactly zero, perhaps due to some yet-to-be-determined symmetry principle -- a 
less arbitrary scenario than 
the SM, in which the value of the CKM phase is a free parameter.

Typically these alternative models for {\em CP} violation lack the distinctive 
feature of the SM that {\em CP} asymmetries are largest in the $B$ sector. 
Many of them can lead to {\em CP} violation in charm decay at the $10^{-3}$ to 
$10^{-2}$ level and have the additional distinctive signatures of large 
flavor-changing neutral currents or mixing in charm.
While direct {\em CP} violation at the $10^{-3}$ level in Cabibbo-suppressed 
charm decays is a prediction of the Standard 
Model~\cite{charm-CP}, its observation in Cabibbo-favored or 
doubly Cabibbo-suppressed decays would constitute unambiguous evidence for new 
physics, as would the observation of indirect {\em CP} violation in charm.
At the levels discussed in the literature,
such effects could be 
detectable in BTeV, which could reconstruct $10^8$ to $10^9$ charm decays, 
but more simulation is 
required to assess backgrounds and systematics~\cite{Strasbourg}.

\section{The BTeV Spectrometer}

\begin{figure}
\vspace{-1.25in}
\centerline{\hspace{-0.8in}\epsfxsize=1.9 in\epsffile{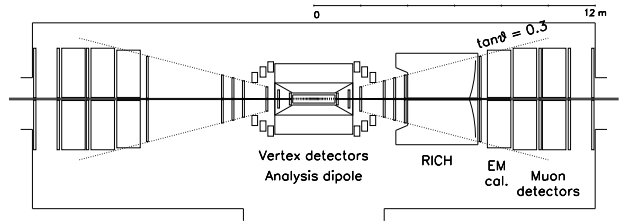}}
\vspace{-0.125 in}
\caption{Sketch of BTeV Spectrometer.\label{fig:spect}}
\end{figure}

The proposed BTeV spectrometer (shown schematically 
in Fig.~\ref{fig:spect}) covers the 
forward and backward regions at the new C0 Tevatron interaction area. The 
instrumented angular range is 
0.01$_\sim$\llap{$^<$}$|\tan{\theta}|_\sim$\llap{$^<$}0.3. Monte Carlo 
simulation shows that such coverage includes $\sim$50\% of all $B$ and $D$ 
decays. 

\begin{figure}
\vspace{-1.2cm}
\centerline{\epsfxsize=3.25 in \epsffile{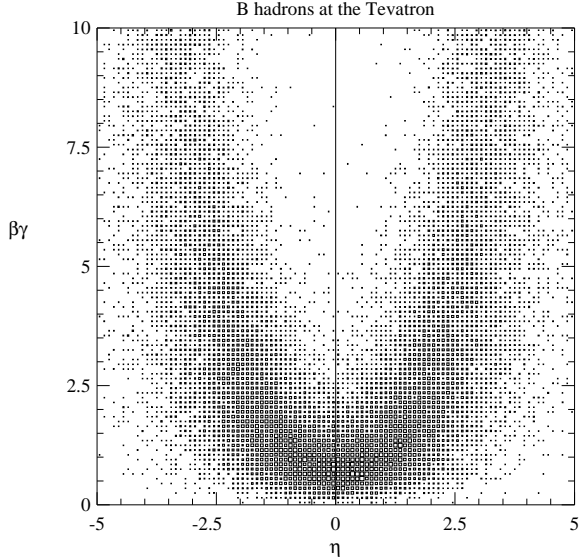}}
\caption{Relativistic boost factor $\beta\gamma$ {\em vs.}\ pseudorapidity 
$\eta$ of $B$ hadrons produced at the Tevatron Collider.\label{fig:bg-vs-y}}
\end{figure}

\begin{figure}
\centerline{\rotatebox{90}{\epsfysize=3.25 in \epsffile{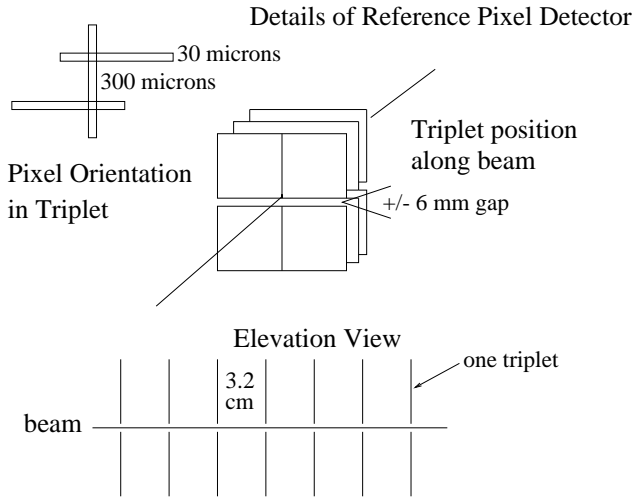}}}
\caption{Proposed arrangement of BTeV vertex detector.\label{fig:vertex}}
\end{figure}

Compared to the ``central-geometry" case (\eg CDF and D0), this 
``forward-geometry" configuration accepts relatively high-momentum particles 
(see Fig.~\ref{fig:bg-vs-y}). 
It also leads to an advantageous vertex-detector arrangement,  consisting of  
detector planes inside the vacuum pipe oriented perpendicular to the beam
(Fig.~\ref{fig:vertex}), 
allowing substantially better reconstruction of decay proper time. Another key 
advantage of forward geometry is the feasibility of effective hadron 
identification. Because QCD mechanisms of $b\bar b$ production yield quark
pairs that are closely correlated in rapidity ($|y_b-y_{\bar
b}|_\sim$\llap{$^<$}1), there is little disadvantage in omitting the
small-rapidity
region: when the decay products of one $B$ hadron are detected in the
forward (or backward) region, decay products of the second (``tagging") $B$
have a high probability to be detected there also.

In addition to large acceptance, the apparatus must have high
interaction-rate capability, 
an efficient trigger for heavy-quark decays, high-speed and 
high-capacity data acquisition,
good mass and vertex resolution, and good particle 
identification. Of these requirements, the most challenging are the 
trigger and the particle identification. We intend to trigger primarily on the
presence of a decay vertex separated from the primary
vertex~\cite{Vertex-trigger}.  To reduce occupancy and facilitate vertex
reconstruction at trigger level 1, pixel detectors will be used for vertex
reconstruction.
For efficient, reliable, and compact particle 
identification, we will build a ring-imaging Cherenkov counter.
In other respects the spectrometer will resemble existing large-aperture
heavy-quark experiments; 
see Refs.~\cite{eoi,Joel-Patty} for more detailed discussions.

\section{Near-Beam Issues in BTeV}

\subsection{Size of vertex-detector beam gap}

\begin{figure}
\centerline{\hspace{2mm}\epsfxsize=3.4 in \epsffile{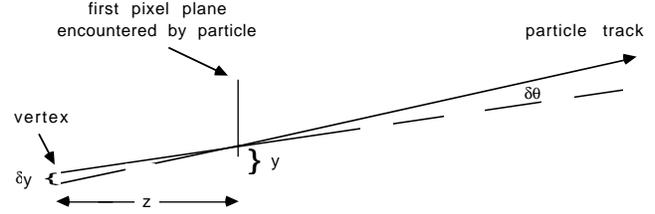}}
\caption{Illustration of dependence of vertical impact-parameter error $\delta 
y$ on scattering angle $\delta\theta$ in first pixel plane.\label{fig:delta-y}}
\end{figure}

A key point in the reconstruction of decay vertices in forward geometry is the 
dependence of the impact-parameter resolution on the transverse distance of 
the vertex detectors from the beam~\cite{Selove}. This is illustrated in   
Fig.~\ref{fig:delta-y}. For sufficiently fine pixel resolution, the 
impact-parameter resolution will typically be dominated by multiple coulomb 
scattering in the first detector plane that the particle encounters. The 
effective r.m.s.\ scattering angle $\delta\theta_y$ in the $y$-$z$ view for a 
charged particle of momentum $p$ traversing a detector of thickness $X$ and 
radiation length $X_0$ is~\cite{PDG}
$$
\delta\theta_y\approx\frac{0.015\,{\rm GeV}}{p}\sqrt{\frac{X}{X_0}}\,.
$$
(The thickness $X$ of course must include substrate, readout electronics, and 
RF shielding.)
If the particle encounters the first detector at a  longitudinal distance $z$ 
from the vertex and transverse distance $y$ from the beam, the scattering 
contribution to impact-parameter resolution is
\begin{eqnarray}
\delta y & \approx & z\delta\theta_y \nonumber \\
& \approx & y\left(\frac{0.015\,{\rm 
GeV}}{p_y}\sqrt{\frac{X}{X_0}}\right)\,.\label{eq:scat}
\end{eqnarray}
A similar equation holds for the $x$-$z$ view, where $\delta x$ also depends 
on $p_y$ since the beam gap is assumed to be in $y$.

We see that the impact-parameter error is proportional to the transverse 
distance of the track from the beam at the first measurement plane encountered 
by the particle. To minimize the scattering contribution, it is thus important 
to keep the beam gap as small as possible. The other parameters appearing in 
Eq.~\ref{eq:scat} are less subject to control by the experimenter: the 
distribution of $p_y$ is determined by the mass and production and decay 
dynamics of the particle to be studied, and $X/X_0$ is fixed by signal/noise, 
mechanical support, and cooling issues. Furthermore, the dependence on $X/X_0$ 
is as the square root, so while thickness should be minimized, it is difficult 
to make a big impact in this way.

\begin{figure}
\vspace{0.75 in}
\centerline{\hspace{-0.4 in}\epsfxsize=3.8 in \epsffile{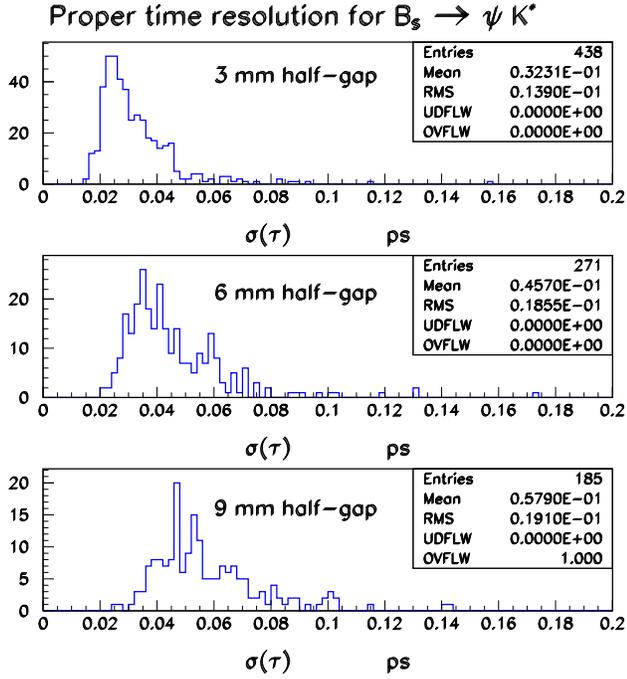}}
\vspace{-1 in}
\caption{Simulated distribution of proper-time resolution for $B_s\to D_s\pi$
events for three different values of $y_{\rm min}$.\label{fig:tres}}
\end{figure}

Fig.~\ref{fig:tres} shows the dependence of the proper-time resolution on the
size of the beam gap for simulated $B_s\to J/\psi K^*$ events. (The time 
resolution in this mode is an indicator of physics reach for studies of $B_s$ 
mixing, a challenging measurement in $b$ physics.)
As the half-gap $y_{\rm min}$ is decreased from 9\,mm to 3\,mm, the r.m.s.\
resolution improves by about a factor of 2. In addition, since cuts on vertex
separation must be made in order to suppress background, the number of events 
in the final sample increases by more than a factor of 2. This indicates the
substantial improvement in physics reach that is possible if the vertex 
detectors can be moved closer to the beam.
With the nominal 6\,mm half-gap, the reach in $x_s$ (the parameter that relates 
the $B_s$ mixing rate to its decay rate) is about 40, {\em i.e.}\ if $x_s=40$ 
we expect to obtain a 5-standard-deviation signal for $B_s$ mixing in about one 
year of running at ${\cal 
L}=5\times10^{31}$\,cm$^{-2}$s$^{-1}$. This should be compared with the 
Standard-Model prediction $x_s<60$ and the current experimental lower limit 
$x_s>15$~\cite{Bs-mixing}.

The size of the half-gap is in principle limited from below by two effects:
1)~radiation damage in the vertex detectors and 2)~creation of backgrounds
at the other interaction regions. In practice the first limit will be reached 
well before the second! For silicon detectors with a 4\,mm half-gap,
the radiation-damage limit ($\sim$10$^{14}$ minimum-ionizing particles/cm$^2$)
is reached in $\approx$1 year
of running at ${\cal L}=5\times10^{31}$\,cm$^{-2}$s$^{-1}$.
Development of diamond pixel detectors may allow a smaller 
gap.\footnote{Subsequently to this Symposium, vertex-detector geometries with a
square beam hole instead of a full-width gap have been simulated and found
to improve physics reach substantially, \eg the $B_s$-mixing reach for one year 
of running at ${\cal L}=5\times10^{31}$\,cm$^{-2}$s$^{-1}$
increases to $x_s\approx 65$.}

\subsection{Wire-target running in C0}

The commissioning of a third collider interaction region is likely to be a
complex process, and simultaneous collider luminosity in all three areas might 
not be available during the first years of Collider Run II.
It has thus been envisaged since the earliest consideration of the C0 
program~\cite{Peoples} that a significant portion of the early running might be 
carried out using a wire or pellet target in the halo of the proton or
antiproton beam. This could afford an early opportunity for commissioning of
detectors. Since it would provide a source of primary interactions localized at
a known point or along a known line in space, it could also be invaluable for
testing the vertex trigger.

While halo running would be essentially useless for beauty due to the small 
fixed-target $b$ cross section~\cite{Jansen}, surprisingly, the charm reach 
could be comparable in fixed-target and collider modes. The increase in charm 
cross section at $\sqrt{s}=2$\,TeV compared to 43\,GeV has not been measured 
but is presumed to be a factor $_\sim$\llap{$^>$}10. However, if only one 
spectrometer arm is instrumented at first, fixed-target has a factor-of-3 
advantage in geometrical acceptance, and a factor $\approx$4 in cross section 
can be made up by taking advantage of the target-$A$ dependence of charm 
production ($\sigma_{c{\bar c}}\propto A^1$~\cite{Leitch} {\it vs.}\
$A^{0.71}$~\cite{PDG} for the
total inelastic cross section which limits the interaction rate). 
Finally, triggering on charm is likely to be considerably more efficient in 
fixed-target mode, where the moderate $p_t$ ($_\sim$\llap{$^<$}1\,GeV) of 
charm decay products stands out more prominently relative to minimum-bias 
background: in fixed-target a factor $\approx$100 in background suppression is 
available {\em before} vertex reconstruction~\cite{eoi}, perhaps allowing 
charm triggering in the short-lifetime regime (proper time $<1$\,ps) crucial 
to studies of charm mixing in $D^0\to$\,hadrons decays~\cite{E791-mixing}.

A possible physics advantage of halo running has also been
suggested~\cite{Kaplan-fcnc}. Biases in charm mixing studies may arise from
$b\to c$ cascade decays. These would be suppressed by two orders of magnitude
in fixed-target relative to collider mode, due to the reduced beauty production
cross section. 

Assuming a 1\,MHz rate of inelastic interactions, $>10^8$ charm decays can be 
reconstructed per year ($10^7$\,s) of fixed-target operation. For example, the 
rate of $D^0 (\overline{D^0})\to K^\mp\pi^\pm$ is estimated 
as~\cite{world-average}
\begin{eqnarray}
n_{D^0 (\overline{D^0})\to K\pi}  =&
10^7 {\rm s} \cdot 10^6 {\rm int./s}  \,\cdot \nonumber \\
&6.5 \times 10^{-4} 
A^{0.29} D^0 ({\overline{D^0}})/{\rm int.} \,\cdot \nonumber \\
& 4\%\cdot10\% \label{eq:charm} \\
=& 1\times10^8\,, \nonumber \hfill\break 
\end{eqnarray}
where the last two factors appearing in Eq.~\ref{eq:charm}
are $B\!R(D^0\to K^-\pi^+)$ and the 
product of acceptance and reconstruction efficiency.
Other decay modes will increase the total by a factor $\sim$3. 
This interaction rate implies $\approx$0.4\,interactions/crossing with 396\,ns 
bunch spacing and $\approx$0.1 with 132\,ns spacing, low enough that 
$p_t$-based triggers should not be badly affected by pile-up. That a 1\,MHz 
interaction rate is feasible with a halo target follows from the work 
of the HERA-$B$ collaboration, who have demonstrated 30\,MHz with wire targets 
in the halo of the HERA proton beam~\cite{HERA-B-tgt}. However, the Tevatron
scraping 
and collimation procedures may need considerable rethinking, since high-rate 
operation of a halo target requires that the target compete efficiently with 
the collimators.

\section{Conclusions}

If approved, BTeV will be the state-of-the-art charm and beauty experiment 
in the mid-2000's period. The
near-beam environment will be key to the experiment's physics reach:
\begin{itemize}
\item
Minimizing the size of the vertex-detector beam gap will both maximize the 
number of events satisfying analysis cuts and optimize their vertex resolution.
\item
Early charm sensitivity at a competitive level may depend on halo targeting.
\end{itemize}

\end{document}